\documentstyle[epsfig]{aipproc}
\begin{document}
\title{Exploring the Small Scale Structure of N103B}
\author{Karen T. Lewis $^1$, David N. Burrows $^1$, John A. Nousek $^1$,\\
Gordon P. Garmire $^1$, Patrick Slane $^2$,   John P. Hughes $^3$}
\address{$^1$ The Pennsylvania State University, 525 Davey Laboratory, University Park, PA 16801 \\
$^2$  Harvard-Smithsonian Center for Astrophysics, 60 Garden Street, Cambridge, MA, 02138 \\
$^3$ Department of Physics and Astronomy, Rutgers University, \\
136 Frelinghuysen Road, Piscataway, NJ 08854-8019}

\maketitle
\begin{abstract}
We present the preliminary results of a 40.8 ks {\it Chandra} ACIS
observation of the young supernova remnant (SNR) N103B located in the
Large Magellanic Cloud. The image reveals structure at the
sub-arcsecond level, including several bright knots and filamentary
structures. The remnant has the characteristic spectrum of a Type Ia
SNR, containing strong lines of Fe, He-and H-like Si and S, Ar, and
Ca. Narrow band images reveal non-uniformities in the remnant.
\end{abstract}
\section*{Introduction}
N103B, one of the brightest X-ray and radio sources in the Large
Magellanic Cloud (LMC), is a young, compact supernova remnant (SNR).
The remnant, located on the north-eastern edge of the HII region N103,
is $\sim$ 40 kpc from the young star cluster NGC 1850. \cite{DM} The
radio and X-ray morphologies are strikingly similar; the western edge
(near the HII region) shows considerable structure and is $\sim$ 3
times brighter than the eastern edge. \cite{DM,Mathewson} In both
bands, the emission arises from a region 7 pc in diameter (d=50kpc to
LMC assumed). Due to its proximity to a star forming region, it was
originally suspected that N103B was the result of the core collapse of
a massive object. However, the ASCA spectrum shows no evidence for
K-shell emission from O, Ne or Mg while Si, S, Ar, Ca, and Fe features
are strong, indicating that the remnant is more likely the result of a
Type Ia SN explosion. \cite{Hughes}
\section*{Observation and Data Cleaning Procedure}
N103B was observed for 40.8 ks with ACIS S-3, one of the
back-illuminated CCDs in the ACIS-S array.  We found that $\sim$35$\%$
of the frames had a high background count rate and that during these
times, the spectrum of the background changed significantly. These
frames were removed from the data set. The standard ASCA grade filter
(g 02346) was used.
\section*{Analysis and Results}
Our observation of N103B has confirmed many earlier results. The {\it
Chandra} X-ray and the Australian Telescope Compact Array (ATCA) radio
images \cite{DM} are quite similar (Figure 1).  Since the radio and
X-ray flux is depressed in the eastern half of the remnant we conclude
that this intensity difference is due to a real density variation, as
opposed to an absorption effect.  While the bright radio knots are in
the general vicinity of the X-ray knots, they do not overlap. Further
investigation is required to determine whether there is a relative
rotation between the two images, which may explain the offset. The
spectrum of N103B as a whole (Figure 2) confirms the earlier ASCA
results. We see no evidence for O or Ne in the spectrum, although a Mg
line is present. 
\begin{figure}[h!]
\centerline{\epsfig{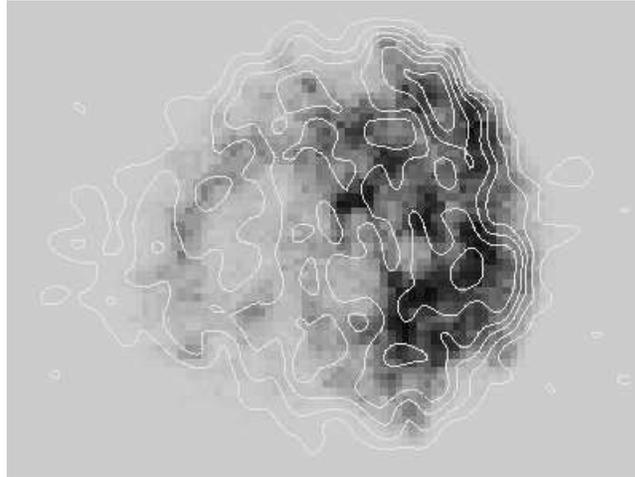}}
\vspace{10pt}
\caption{Plot of X-ray image with 2.5 cm ATCA contours [1]. Contours are 1.7, 
1.3, 0.9, 0.6, 0.3, and 0.1 mJy/beam. The spatial resolution of the radio data is 1.2'',  which is slightly 
larger than the Chandra's. }
\label{fig1}
\end{figure}
\begin{figure} 
\centerline{\epsfig{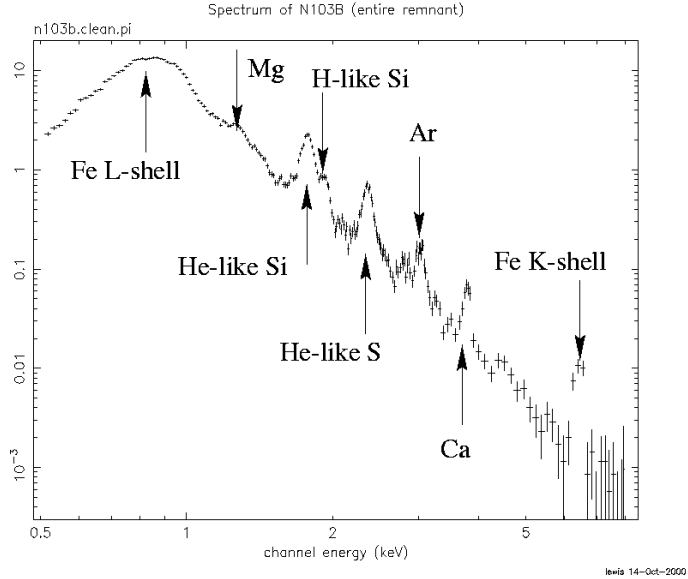}}
\vspace{10pt}
\caption{Spectrum of N103B as a whole. The data are plotted so that each data point has a 6 sigma 
significance with a maximum bin size of $\sim$ 65eV}
\label{fig2}
\end{figure}

\indent Continuum subtracted (CS) and equivalent width (EW)
images show significant variation throughout the remnant.  The images
for He-like Si are shown in Figure 3.  The interval 1700-1865 eV was
used to create the He-like Si line image.  To estimate the continuum,
we performed a linear interpolation between images in the 1515-1680 eV
and 2050-2215 eV bands, which are free of any significant lines.  All
images were smoothed with a 3 pixel (1.5'') boxcar to reduce the
number of pixels with no counts.  We were concerned that the CS and EW
images were unreliable, as each pixel contained only a few counts at
most, so the CS and EW images were collapsed into radial and azimuthal
dimensions by binning the line and continuum images into rings and
sectors.  The CS and EW flux were calculated for each ring and sector
as shown in bottom of Figure 3.  The radial and azimuthal plots
confirm what is seen in the images. The emission from He-like Si is
greatest along the western edge, where the entire remnant is
bright. However, as shown by the EW images and plots, it is only in
the outer rim of the remnant that the He-like Si emission drastically
dominates over the continuum emission. Throughout the remainder of the
remnant, the ratio of line to continuum is relatively flat.  We have
not yet determined whether the change in emissivity is due
to a difference in abundance, temperature, ionization or a combination
of these.
\section*{Conclusions}
Preliminary analysis of this {\it Chandra} observation reveal that
N103B shows significant structure at the arcsecond level. The
variations in X-ray flux are consistent with those in radio and are
due to real density variation within the remnant. The spectrum of the
remnant is that of a typical Type Ia SNR, however future spatially
resolved spectroscopy will allow for a better classification.
Finally, there are dramatic spatial variations in the emissivity of
He-like Si in particular. The radial variation in the equivalent width 
of the remnant is particularly intriguing. Detailed spectral analysis of 
concentric rings will enable us to determine the source of these 
and other similar variations.

\begin{figure}[t!]
\centerline{\epsfig{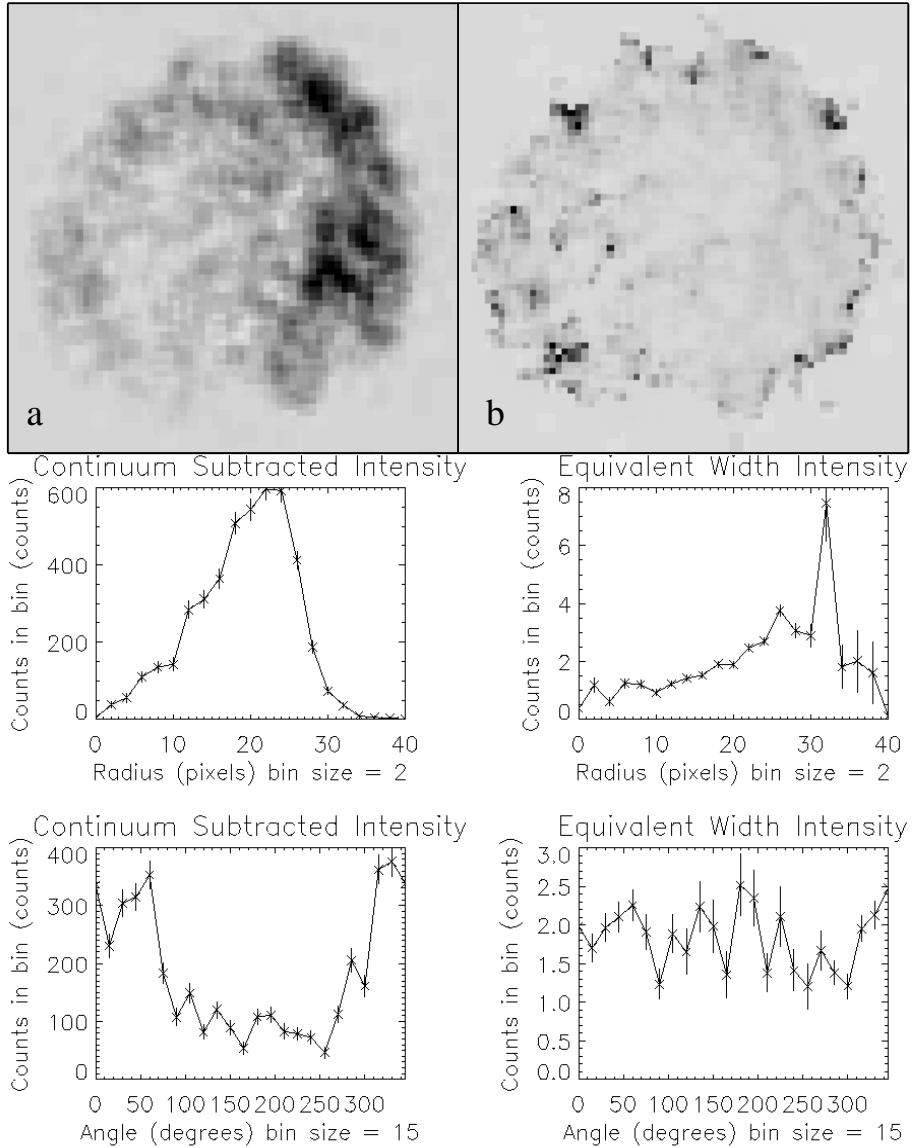}}
\vspace{10pt}
\caption{Plot of He-like Si continuum subtracted (a) and equivalent width images (b).  Below are 
radial and azimuthal plots. 2 pixels = 1'' $\sim$ 0.2 pc}
\label{fig3}
\end{figure}

\end{document}